\documentclass[%
reprint,
superscriptaddress,
 amsmath,amssymb,
 aps,
prl,
longbibliography,
floatfix,
]{revtex4-2}
\usepackage{graphicx}
\usepackage{color}
\usepackage{subfig}
\usepackage{hyperref}
\usepackage{float}
\usepackage{array, multirow}
\usepackage{amsmath}
\usepackage{newtxtext}      \usepackage{booktabs}
\usepackage{xcolor}
\usepackage{soul}
\usepackage{url}
\usepackage{qcircuit}
\usepackage[margin=0.8in]{geometry}

\usepackage{titlesec}
\titleformat{\section}
  {\normalfont\fontsize{12}{15}\bfseries}{\thesection}{1em}{}

\hypersetup{colorlinks,
    linkcolor={black!75!black!80!yellow},
    citecolor={black!75!black!80!yellow},
    urlcolor={black!75!black!80!yellow}
    }

\begin{document}


\title{Reducing the Quantum Many-electron Problem to Two Electrons with Machine Learning}

\author{LeeAnn M. Sager-Smith}
\affiliation{Department of Chemistry and The James Franck Institute, The University of Chicago, Chicago, IL 60637 USA}
\author{David A. Mazziotti}
\email{damazz@uchicago.edu}
\affiliation{Department of Chemistry and The James Franck Institute, The University of Chicago, Chicago, IL 60637 USA}

\date{Submitted July 6, 2022\color{black}; Revised August 19, 2022\color{black}}



\begin{abstract}

\noindent\textbf{Abstract.} An outstanding challenge in chemical computation is the many-electron problem where computational methodologies scale prohibitively with system size. The energy of any molecule can be expressed as a weighted sum of the energies of two-electron wave functions that are computable from only a two-electron calculation.  Despite the physical elegance of this extended ``aufbau'' principle, the determination of the distribution of weights\textemdash geminal occupations\textemdash for general molecular systems has remained elusive.  Here we introduce a new paradigm for electronic structure where approximate geminal-occupation distributions are ``learned'' via a convolutional neural network.  We show that the neural network learns the $N$-representability conditions, constraints on the distribution for it to represent an $N$-electron system.  By training on hydrocarbon isomers with only 2-7 carbon atoms, we are able to predict the energies for isomers of octane as well as hydrocarbons with 8-15 carbons.  The present work demonstrates that machine learning can be used to reduce the many-electron problem to an effective two-electron problem, opening new opportunities for accurately predicting electronic structure.

\end{abstract}

\maketitle

\section{Introduction}
\vspace{-5mm}

\noindent For any molecular system, the Schr\"odinger equation can, \textit{in theory}, be solved exactly using a full configuration interaction (FCI) calculation \cite{Knowles1989,Jensen2017,Bauschlicher2007} with a complete basis set; however, \textit{in practice}, the computational complexity of such an exact approach grows factorially with system size \cite{Bauschlicher2007}, making molecular systems with more than a few dozen electrons intractable.  Over time, many approximate methodologies have been introduced in an attempt to obtain ``good enough'' solutions to the electronic Schr\"odinger equation that predict energies within chemical accuracy ($\thicksim$1 kcal/mol).

Hartree-Fock theory\textemdash a mean-field approach\textemdash yields reasonable results for a wide array of molecular systems containing up to a few hundred atoms~\cite{Jensen2017}; however, it fails in molecules in which the motions of electrons are significantly correlated.  Techniques which more-accurately capture correlation energy such as many-body perturbation theory, coupled cluster theory, complete active-space self-consistent field theory, and others remain computationally expensive for large system sizes \cite{Mezey2011,Jensen2017}.  The so-called many-electron problem\textemdash whereby the cost of highly-accurate \textit{ab initio} computational methodologies scales in a prohibitive manner with system size\textemdash is hence an outstanding challenge in chemical computations.

Machine learning may enable us to circumvent this problem by allowing us to use information about smaller molecules to treat correlation in larger systems at a reduced cost \color{black} \cite{Sajjan_2022}\color{black}.   It has been used to learn the energies of various molecular structures~\cite{Sureshbabu2021, PhysRevB.101.195141, Torlai2018, ChNg2017}, new functionals for density functional theory (DFT)~\cite{Brockherde2017,PhysRevLett.108.253002,Gedeon2021},  \color{black} inverse problems in electronic structure theory~\cite{Mills_2017,Wang_2020}, \color{black} and  even  the many-body wave function of one-dimensional spin systems~\cite{carleo2017solving}.  However, these areas are in their early stages and have yet to demonstrate definite success in decreasing the degree of scaling with system size.

\begin{figure*}[tb!]
    \hspace{-0.75cm}
    \centering
    \includegraphics[width=18cm]{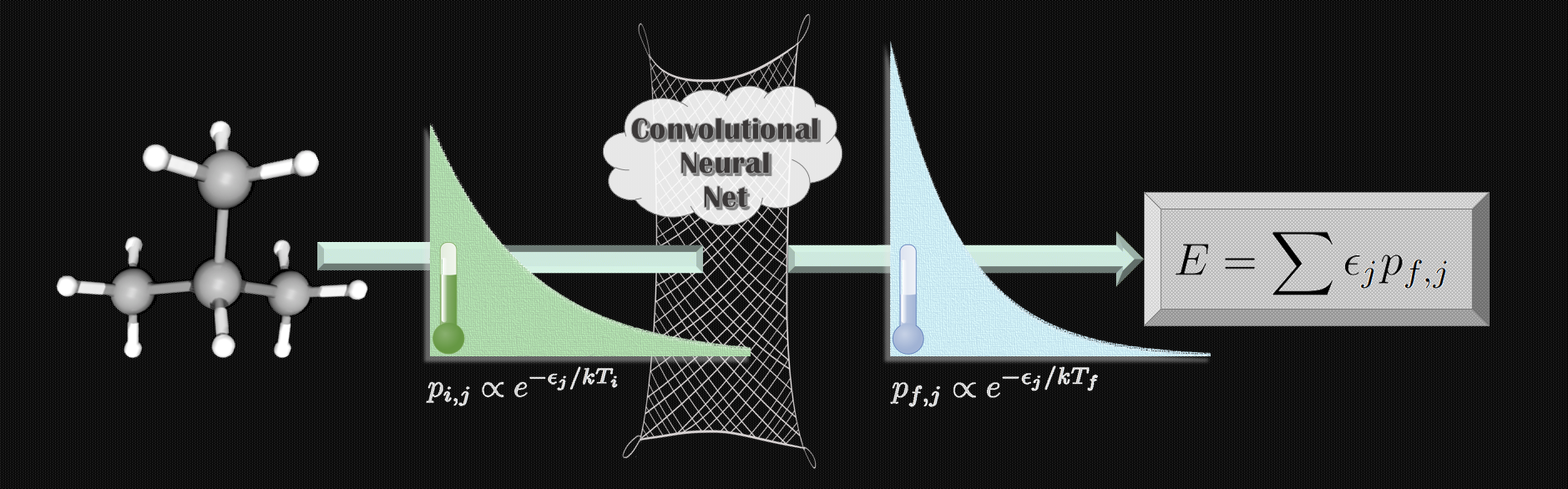}
    \caption{\textbf{Graphic demonstrating algorithm flow.}  For a given molecule, a trained convolutional neural network is used to predict the Boltzmann-like correlation temperature ($T_f$) with the eigenfunctions of the reduced Hamiltonian ($\epsilon_j$) and the Hartree-Fock correlation temperature ($T_i$) as inputs.  The correlation temperature ($T_f$) allows for the approximation of the geminal populations ($p_{f,j}$) by Eq. (\ref{eq:distribution}), which 
    is sufficient for the prediction of the energy by Eq. (\ref{eq:BoppE}).}
    \label{fig:inforgraphic}
\end{figure*}

In this \emph{Article} we introduce a new paradigm for utilizing machine learning in quantum chemistry in which we reduce the quantum many-electron problem to a more tractable, better scaling two-electron problem.  As originally proposed by Bopp~\cite{Coleman2000,Haar1961}, the energy of a molecule of arbitrary size can be expressed without approximation as a weighted sum of the energies of two-electron wave functions, known as geminals.  However, despite its physical significance as an extension of the ``aufbau'' principle, the distribution of weights\textemdash geminal occupations\textemdash has remained elusive.   Here, we show that the geminal-occupation distribution can be learned with machine learning.  We use a convolutional neural network (CNN) to learn an effective temperature in a Boltzmann-like distribution for the geminal occupations.  The effective temperature\textemdash or correlation temperature\textemdash is inversely related to  the electron correlation.  The neural network, we demonstrate, learns the $N$-representability of the distribution\textemdash the representability of the distribution by an $N$-electron system~\cite{Mazziotti2016a, Shenvi2010, piris_2017, C1963}, which appears as a nonzero temperature.  The scheme can be viewed as a two-electron reduced density matrix (2-RDM) theory as the geminal occupations are an integral part of the 2-RDM.  A schematic of the machine learning algorithm for predicting molecular energies is shown in Fig.~\ref{fig:inforgraphic}.


We apply the machine learning algorithm to hydrocarbon systems.  Specifically, by training a convolutional neural network on all isomers of ethane through heptane, we predict the correlation temperatures\textemdash and hence molecular energies\textemdash of all of the isomers of octane as well as all straight-chained hydrocarbons from octane through pentadecane.  We find that this RDM-based machine learning method accurately recovers the correlation energy for larger hydrocarbon systems, with the $N$-representability conditions being learned by the CNN framework.  Our approach\textemdash which scales as $O[n^6]$\textemdash improves upon the exponential scaling of traditional configuration-interaction calculations, foreshadowing the potential utility of this machine-learning reduced density matrix approach to the determination of accurate molecular energies.  \color{black} While polynomial-scaling levels of theory such as Coupled Cluster with Single and Double Excitations (CCSD) can be used to treat weakly-correlated systems such as the hydrocarbons presented in this manuscript, if trained on appropriate molecular data, our convolutional network approach may be capable of accurately recovering correlation energy for more highly-correlated systems.   \color{black}

\section{Results and Discussion}
\vspace{-5mm}

\noindent\textbf{Theory.} Central to our modern understanding of chemistry is the concept of the molecular orbital.  Any molecule's electronic structure can be readily understood in terms of its molecular orbitals which are filled from lowest-in-energy to highest-in-energy by the Pauli exclusion principle.  When electrons of a molecule become strongly correlated, however, the orbital picture with unit filling of the lowest orbitals breaks down.  Because electronic interactions are at most pairwise, the orbital picture can in principle be replaced by an exact two-electron (geminal) picture, which is derivable from 2-RDM theory.

The ground- or excited-state energy of any atom or molecule is expressible as an exact functional of the 2-RDM (${}^{2}D$) \cite{RDM2007, Coleman2000, Coleman1963, Garrod1964, Erdahl1978, Erdahl1979, Erdahl1989, Mazziotti1998b, Mazziotti1999a, Erdahl2000, Nakata2001, Mazziotti2002b, M2004, Zhao2004, Gidofalvi2005, Mazziotti2006c, Mazziotti2006, Cances2006, Mazziotti2007d, Erdahl2007, Kais2007, Gidofalvi2008, Mazziotti2008a, Gidofalvi2009, Snyder2010, Shenvi2010, Mazziotti2011, Mazziotti2012e, Mazziotti2012b, Verstichel2012, Sand2015, Poelmans2015, Schlimgen2016, Mazziotti2016a, Mazziotti2016, Hemmatiyan2018, Alcoba2018, Boyn2020, Xie2020, Boyn2021a, Ewing2021}
\begin{equation}
E = \int{ ^{2} {\hat K} \, ^{2} D({\bar 1}{\bar 2};12) d1 d2 }
\end{equation}
where $^{2} {\hat K}$ is the reduced Hamiltonian operator
\begin{gather}
^{2} {\hat K} = -\frac{N}{2} \left ( \frac{{\hat p}_{1}^{2}}{2m} + \frac{{\hat p}_{2}^{2}}{2m}  + \sum_{k}{ \frac{Z_{k}}{r_{1k}} } + \sum_{k}{ \frac{Z_{k}}{r_{2k}} } \right )\nonumber \\+ \frac{N(N-1)}{2} \frac{1}{r_{12}} .
\label{eq:redHam}
\end{gather}

\begin{figure*}[tb!]
    \centering
    \includegraphics[width=16cm]{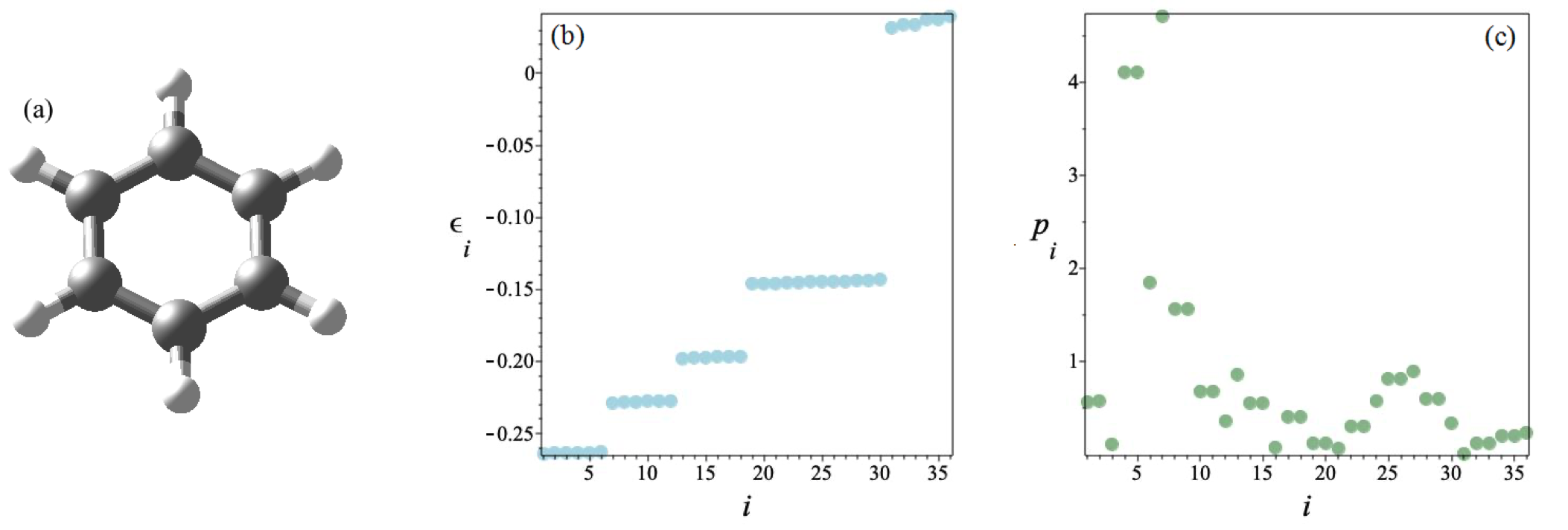}
    \caption{\textbf{Example of geminal energies and probabilities.} For (a) benzene, we can use the (b) geminal energies $\epsilon_i$ to learn the (c) geminal probabilities $p_i$\textemdash both of which are computed here from a [$N_e=6, \ N_o=6]$ complete active-space self-consistent-field (CASSCF) using the minimal Slater-type orbital basis set with six Gaussian primitive functions representing each Slater-type orbital (STO-6G).  Knowing both geminal energies and geminal populations is sufficient to determine molecular energies via Eq. (\ref{eq:BoppE}).}
    \label{fig:benzene}
\end{figure*}

In a finite orbital basis set, the operators are expressible as a reduced Hamiltonian matrix.   Diagonalization of this reduced Hamiltonian matrix yields a set of eigenvalues and eigenvectors (or geminals).  In the basis set of geminals, the Hamiltonian is a diagonal matrix consisting of its eigenvalues, the 2-RDM has a non-negative diagonal elements which we denote by $p_{i}$, and  energy is the sum over the geminal eigenvalues of the Hamiltonian matrix $\epsilon_{i}$ weighted by the non-negative geminal occupations $p_{i}$:
\begin{equation}
    E = \sum_{i}{ p_{i} \epsilon_{i} }.
    \label{eq:BoppE}
\end{equation}  By this transformation we express the energy as a functional of the eigenvalues of the reduced Hamiltonian $\epsilon_{i}$, which are readily computed at the cost of the two-electron calculation, and the unknown geminal occupations $p_{i}$ (see Fig. \ref{fig:benzene}).

The German chemist Bopp originally proposed approximating the geminal occupation numbers by a Pauli-like filling scheme \cite{Coleman2000,Haar1961}.  He suggested choosing the lowest $N(N-1)/2$ to be equal to one.  This approach, while analogous to the filling of orbitals in molecular-orbital theory, generates accurate energies for four-electron atoms and ions but energies for larger molecular systems that are too low.  Coleman suggested that the filling of the geminal by two electrons\textemdash or the pseudo-particle called a pairon\textemdash should follow a fundamental probability distribution as in statistical mechanics \cite{Coleman2000}.  He proposed a Boltzmann distribution
for the geminal occupations based on the geminal energies.
While such a distribution is not exact because the pairon pseudo-particles obey neither the Fermi-Dirac or Bose-Einstein particle statistics, there exists a Boltzmann-like distribution given by
\begin{equation}
    p_i = \frac{N(N-1)}{Z}e^{-\epsilon_i/kT^*}
    \label{eq:distribution}
\end{equation}
and parameterized by a specific correlation temperature ($T^*$) such that the resultant approximate geminal probability distribution allows for the accurate computation of a molecule's energy according to Eq. (\ref{eq:BoppE}).  However, the ability to determine such a correlation temperature is currently only possible if the geminal energies ($\epsilon_i$) and geminal populations ($p_i$) are both known.

Here, we train a convolutional neural network (CNN) to predict the correlation temperature for a given molecular system consistent with its ground-state energy.  The convolutional neural network is trained on inputs corresponding to both geminal energies\textemdash expressed as partition functions given by
\begin{equation}
    Z=\sum\limits_i e^{-\epsilon_i/kT}
    \label{eq:partition}
\end{equation}
for a variety of temperatures\textemdash as well as the computed Hartree-Fock correlation temperature ($T^*_{HF}$) and with training outputs corresponding to a $\Delta$ value representing the difference between the exact (i.e. configuration interaction) correlation temperature and the HF correlation temperature, i.e., $\Delta = T^*_{EXACT}-T^*_{HF}$.  For larger molecular systems, we then predict the $\Delta$ values by reading in the geminal energies and Hartree-Fock correlation temperatures for those molecules into the trained neural network.  These $\Delta$ values are then added to the $T_{HF}^*$s in order to yield the exact correlation temperatures, which allows for the approximation of the geminal probability distributions and hence the molecular energies via Eq. (\ref{eq:BoppE}).

In general, for two-electron reduced density matrix methodologies, the 2-RDM must be constrained to represent the $N$-electron wavefunction through application of $N$-representability constraints \cite{Mazziotti2016a,Shenvi2010,piris_2017,C1963}.  Here, if $N$-representability conditions are not accounted for in our Boltzmann-like machine learning approach, the correlation temperature would be zero, which corresponds to the lowest-energy geminal being fully occupied by all electron pairs.   This electronic structure machine learning approach, however, maintains $N$-representability by learning correlation temperatures from $N$-representable training data and applying this inherent ``learned'' $N$-representability to the testing data.

See the \color{black} Experimental \color{black} section \color{black} at the end of this document \color{black} for additional details.

\begin{figure*}
    \begin{center}
    \subfloat[\color{black}CASSCF, cc-pVDZ, {[}$N_e=8, \ N_o=8${]} \color{black}]{\includegraphics[width=15cm]{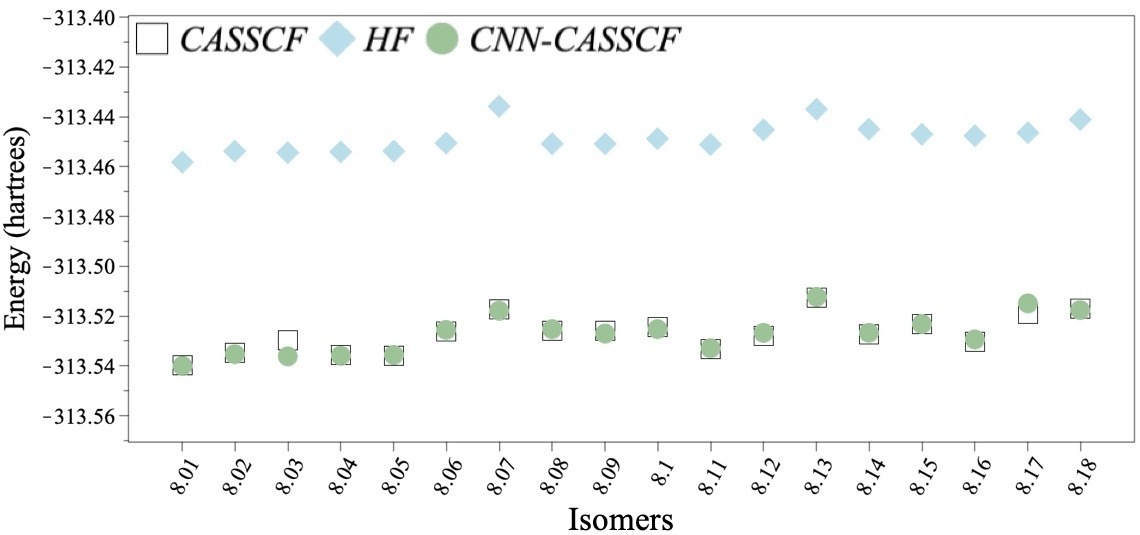}}\\
    \subfloat[\color{black}CCSD, cc-pvDZ\color{black}]{\includegraphics[width=15cm]{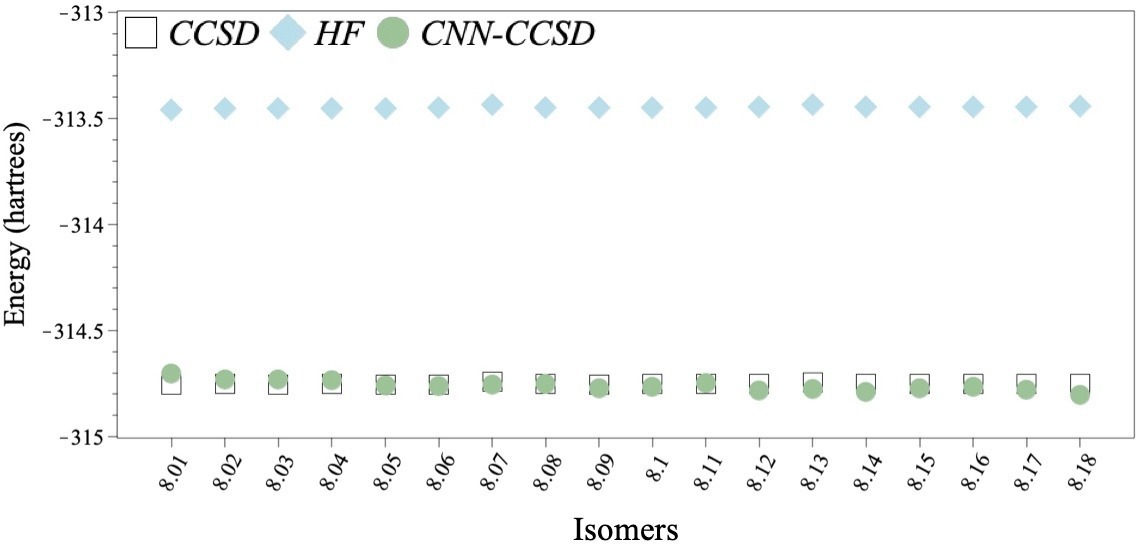}}
    \end{center}
    \caption{\textbf{Octane data.} Hartree-Fock energies (HF, blue diamonds), \color{black} (a) Complete Active Space Self-Consistent Field/(b) Coupled Cluster Single Double (CASSCF/CCSD, black boxes) energies, \color{black} and energy values predicted via utilization of Convolutional Neural Networks (CNN, green circles) are shown for the series of octane isomers.  As can be seen, the CNN methodology trained on smaller hydrocarbon data fairly accurately recovers the correlation energy. Isomer labels are given by [8.01: `Octane', 8.02: `2-Methylheptane', 8.03: `3-Methylheptane', 8.04: `4-Methylheptane', 8.05: `2,2-Dimethylhexane', 8.06: `2,3-Dimethylhexane', 8.07: `2,4-Dimethylhexane', 8.08: `2,5-Dimethylhexane', 8.09: `3,3-Dimethylhexane', 8.10: `3,4-Dimethylhexane', 8.11: `3-Ethylhexane', 8.12: `2,2,3-Trimethylpentane', 8.13: `2,2,4-Trimethylpentane', 8.14: `2,3,3-Trimethylpentane', 8.15: `2,3,4-Trimethylpentane', 8.16: `3-Ethyl-2-Methylpentane', 8.17: `3-Ethyl-3-Methylpentane', 8.18: `2,2,4,4-Tetramethylbutane']. \color{black} Hartree-Fock, CASSCF, and CCSD  calculations are all \color{black} computed here using Dunning's double-zeta (cc-pVDZ) basis set with the CASSCF calculations employing a [$N_e=8, \ N_o=8$] active space.}
    \label{fig:octane}
\end{figure*}

\vspace{3mm}
\noindent\textbf{Energetic Predictions for Isomers of Octane.} For the eighteen isomers of octane\textemdash with molecular geometries obtained from the PubChem database \cite{PubChem_2021}\textemdash, the Hartree-Fock and CASSCF energies are computed using Dunning's double-zeta (cc-pVDZ) basis set with complete active-space self-consistent-field (CASSCF) calculations employing a [$N_e=8, \ N_o=8$] active space.  Utilizing a convolutional neural network trained on hydrocarbons ranging from two to seven carbon atoms, the correlation temperature corresponding to the CASSCF energy is predicted for each of the octane isomers and used to compute the predicted CASSCF energies shown in Fig. \ref{fig:octane}(a).  As can be seen from this figure, which shows energy versus isomer identifier, the predicted CASSCF energies (green circles) show good agreement with the actual CASSCF energies (black boxes), vastly improving upon the Hartree-Fock energies (blue diamonds), and hence our predictions capture the correlation energy in a fairly accurate manner.

\color{black}
Additionally, in order to demonstrate the generality of our reduced density matrix approach for ``learning'' molecular energies, Coupled Cluster Single Double (CCSD) energies are computed for the cc-pVDZ basis for hydrocarbons ranging from two to seven carbon atoms.  The corresponding CCSD correlation temperatures are then used to train a convolutional neural net, and the correlation temperature corresponding to the CCSD energy is then predicted for each isomer of octane, with the resultant predicted CCSD energies shown in Fig. \ref{fig:octane}(b).  Similar to the CASSCF energies from Fig. \ref{fig:octane}(a), the CCSD predicted energies (green circles) demonstrate good agreement with the actual CCSD energies (black boxes) when compared to the Hartree-Fock energies (blue diamonds).  Hence, for this second level of theory, our predictions capture correlation energies in a fairly accurate manner.  Additional predictions corresponding to CCSD calculations utilizing the STO-6G basis set can be seen in the Supporting Information.
\color{black}

We next explore systems composed of larger hydrocarbons to determine whether such good agreement remains consistent as system size is increased while the training data remains the same.

\begin{figure*}
    \centering
    \includegraphics[width=10cm]{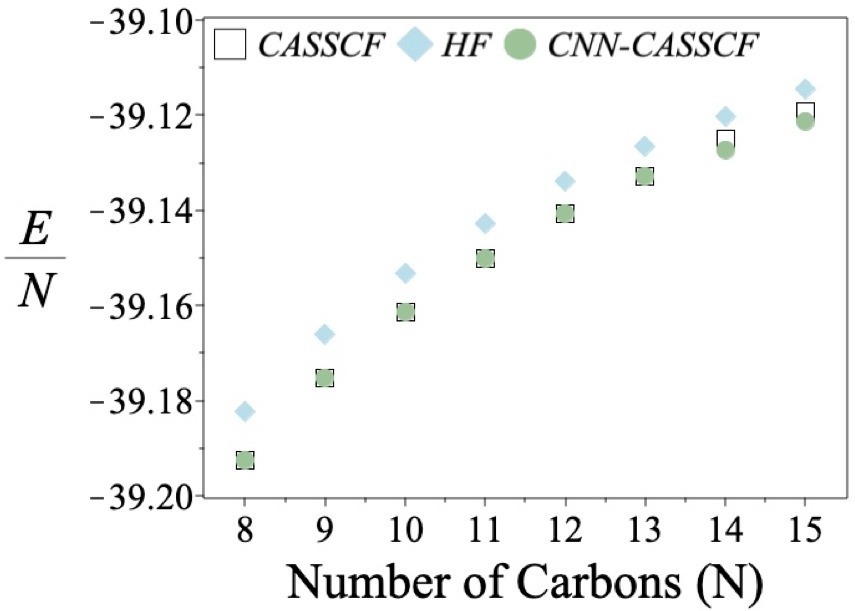}
    \caption{\textbf{Large hydrocarbon data.} Hartree-Fock energies (HF, blue diamonds), Complete Active Space Self-Consistent Field energies (CASSCF, black boxes), and energy values predicted via utilization of Convolutional Neural Networks (CNN, green circles) per number of carbons are shown for the series of straight-chained hydrocarbons from octane through pentadecane.  As can be seen, the CNN methodology trained on smaller hydrocarbon data fairly accurately recovers the correlation energy.  Isomer labels are given by [8: `Octane', 9: `Nonane', 10: `Decane', 11: `Undecane', 12: `Dodecane', 13: `Tridecane', 14: `Tetradecane', 15: `Pentadecane']. Both Hartree-Fock and CASSCF calculations are computed here using Dunning's double-zeta (cc-pVDZ) basis set with the CASSCF calculations employing a [$N_e=8, \ N_o=8$] active space.}
    \label{fig:isomers}
\end{figure*}

\vspace{3mm}
\noindent\textbf{Energetic Predictions for Large Hydrocarbon.} For the eight straight-chained hydrocarbons ranging from octane to pentadecane\textemdash with molecular geometries obtained from the PubChem database \cite{PubChem_2021}\textemdash, the Hartree-Fock and CASSCF energies are computed using Dunning's double-zeta (cc-pVDZ) basis set with the CASSCF calculations employing a [$N_e=8, \ N_o=8$] active space.   Utilizing a convolutional neural network trained on hydrocarbons ranging from two to seven carbon atoms, the correlation temperature corresponding to the CASSCF energy is predicted for each of the octane to pentadecane hydrocarbon isomers and used to compute the predicted CASSCF energies shown in Fig.~\ref{fig:isomers}.  As can be seen from this figure, which shows energy per carbon versus number of carbons, the predicted CASSCF energies (green circles) show good agreement with the actual CASSCF energies (black boxes), vastly improving upon the Hartree-Fock energies (blue diamonds), and hence our predictions capture the correlation energy in a fairly accurate manner.  Although there is a slight increase in the error as system size is increased, it appears to be small enough that the energies of even larger hydrocarbon isomers may be able to be predicted in an accurate manner through use of our convolutional neural network trained on only hydrocarbons with seven or fewer carbon atoms.  \color{black} Similar promising results are obtained for predicting CASSCF energies for octane, nonane, decane, and undecane via a convolutional neural network trained on CASSCF calculations for hydrocarbons with two to seven carbons that utilize a [$10$,$10$] active space and the cc-pVTZ basis set as can be seen in the Supporting Information.\color{black}

\section{Conclusions}
\vspace{-5mm}
\noindent In this \textit{Article}, we introduce a new paradigm based on a two-electron, reduced density matrix approach for the utilization of machine learning architecture in the prediction of accurate correlation energies for molecular systems at reduced computational expense.  By employing a Boltzmann-like distribution for two-electron geminal populations parameterized by a correlation temperature, we train a convolutional neural network on correlation temperatures corresponding to CASSCF \color{black} and CCSD \color{black} calculations for smaller molecular systems in order to predict CASSCF \color{black} and CCSD \color{black} correlation temperatures for larger, more computationally-expensive molecular systems and hence obtain predicted CASSCF\color{black}/CCSD \color{black} energies.  Moreover, the $N$-representability conditions are inherently maintained by our CNN framework\textemdash as evinced by nonzero correlation temperatures.  This methodology for the prediction of CASSCF energies scales as $O[n^6]$ with the number of orbitals due to the diagonalization of the reduced Hamiltonian, which is an improvement over the exponential scaling of a traditional CASSCF calculation.  \color{black} See the Experimental section for additional comments on computational scaling. \color{black}

Demonstrating the power of this technique, we train a convolutional neural network on small hydrocarbon systems\textemdash with the number of carbon atoms ranging from two to seven\textemdash in order to predict CASSCF energies for larger hydrocarbon systems\textemdash with the number of carbons ranging from eight to fifteen.  We find that our RDM-based machine learning approach accurately recovers the correlation energy for the larger hydrocarbon systems.  Thus, our trained convolutional neural network allows us to predict CASSCF-like results at significantly lower computational expense.

\color{black}
While the hydrocarbons involved in training and testing this implementation of our machine-learning reduced density matrix approach do not demonstrate large degrees of correlation, the prediction of accurate correlation energies for larger molecular systems of the type included in the training set likely indicates that as long as the convolutional neural network is trained on appropriate small molecules, the energies of highly-correlated, larger molecules should be able to be obtained via our methodology.  Specifically, if one wishes to predict the energy of a molecule which demonstrates a fairly-large degree of correlation, smaller correlated systems would likely be necessary to train the neural network.  Application of our machine-learning reduced density matrix approach to highly-correlated systems is a future direction of this research.
\color{black}

This work foreshadows the promise of machine learning in molecular electronic structure calculations, demonstrating that ``learning'' information about less-expensive, smaller molecular systems can be directly applied to larger typically more-expensive molecules. Future electronic structure methodologies may even include pre-trained convolutional neural networks\textemdash possibly varying with the types of atoms, basis set, active space, functional groups, and/or degree of bond saturation inherent to the molecular system of interest\textemdash trained on FCI (or similarly expensive) correlation temperatures.  This work serves as an initial step in the realization of a combined reduced-density-matrix and machine-learning approach that may provide a real advance in decreasing computational expense for large, highly-correlated electronic structure calculations.

\color{black}
\section{Supporting Information}
\vspace{-5mm}
\noindent An analysis of the effect of changing the active space size on our machine-learning reduced density matrix approach; application of our reduced density matrix machine learning algorithm to the prediction of CASSCF energies with a [$10$,$10$] active space and cc-pVTZ basis set; application of our reduced density matrix machine learning algorithm to the prediction of CCSD energies with a STO-6G basis.
\color{black}

\section{Acknowledgements}
\vspace{-5mm}
\noindent D.A.M. gratefully acknowledges support from U.S. National Science Foundation under Grant No. 2155082 and the Department of Energy, Office of Basic Energy Sciences under Grant No. DE-SC0019215.  L.M.S.-S. also acknowledges support from the U. S. National Science Foundation under Grant No. DGE-1746045.


\section{References}
\vspace{-20mm}
\bibliography{references,references_2,prop}

\section{Experimental}
\vspace{-5mm}
\noindent \textbf{Computational Methods.} The molecular geometries for all hydrocarbon isomers are obtained from the PubChem database \cite{PubChem_2021}.  Molecular energies are then computed \color{black} for Hartree-Fock, Complete Active Space Self-Consistent Field (CASSCF), and Coupled Cluster Single Double (CCSD) \color{black} levels of theory through use of a Dunning's double-zeta (cc-pVDZ) basis set, with the CASSCF calculations employing a [$N_e=8, \ N_o=8$] active space.  These calculations are accomplished via the Quantum Chemistry Toolbox \cite{Montgomery2020,QCP_2022} in the Maple computing environment \cite{maple_2022}.  Note that while\textemdash throughout this text\textemdash the size of the active space for the training and testing molecules is made identically [$N_e=8, \ N_o=8$] for all CASSCF calculations, changing active space sizes with the number of carbons yielded similar results to those we present here.  (See the Supporting Information for additional details.)

\vspace{3mm}
\noindent \textbf{Computation of Geminal Energies and Populations.} The reduced Hamiltonian (${}^{2}K$) shown in Eq. (\ref{eq:redHam}) is obtained by directly computing the one electron integrals and the electron repulsion integrals via the MOIntegrals function of the Quantum Chemistry Toolbox \cite{Montgomery2020} in the Maple computing environment \cite{maple_2022} and then applying the appropriate conversions to put it into the same orbital basis as the 2-RDM.  The geminal energies ($\epsilon_i$) then correspond to the eigenvalues of the ${}^{2}K$ matrix.  The populations ($p_i$) of the geminals are then obtained via the following
\begin{equation}
    p_i=\langle v_i | {}^{2} D | v_i \rangle
    \label{eq:pop_from_v}
\end{equation}
where $v_i$ is the eigenvector of the reduced Hamiltonian corresponding to the the geminal energy $\epsilon_i$ and where ${}^{2}D$ is the particle-particle reduced density matrix (2-RDM).

\vspace{3mm}
\noindent \textbf{Convolutional Neural Network.}

\noindent \textit{Model Inputs.} For a given molecular system,  both the geminal energies ($\epsilon_i$) and the Hartree-Fock correlation temperature ($T_{HF}^*$) are input into the convolutional neural network.  Specifically, the geminal energies are encoded as partition functions ($Z$)\textemdash computed according to Eq. (\ref{eq:partition})\textemdash for $\beta$ values ranging from 0 to 20 by 0.4 where
\begin{equation}
    \beta=\frac{1}{k T}
\end{equation}
and where $k$ is the Boltzmann constant.  The Hartree-Fock correlation temperature is obtained by inserting Eq. (\ref{eq:partition}) into Eq. (\ref{eq:distribution}) which is inserted into Eq. (\ref{eq:BoppE}) to obtain
\begin{equation}
    E(T)=\frac{N(N-1)}{\sum\limits_i e^{i\epsilon_i/kT}}\sum\limits_j \epsilon_j e^{-\epsilon_j/kT}
    \label{eq:opteq}
\end{equation}
and then temperature is optimized via scipy.optimize.minimize such that $|E_{HF}-E(T)|$ is minimized.

\vspace{2mm}
\noindent \textit{Model Outputs.} For a given molecular system, the output of the convolutional neural net is a $\Delta$ value representing the difference between the Hartree-Fock correlation temperature and the predicted CASSCF correlation temperature, i.e., $\Delta = T^*_{CAS}-T^*_{HF}$.  From this output, the predicted correlation temperature corresponding to the CASSCF calculation can be computed by adding the output ($\Delta$) to the Hartree-Fock correlation temperature ($T^*_{HF}$), which can be used\textemdash along with the known geminal energies ($\epsilon_i$)\textemdash to calculate the predicted CASSCF energy according to Eq. (\ref{eq:opteq}).

\vspace{2mm}
\noindent \textit{Training Data.} All hydrocarbons isomers ranging from two to seven carbon atoms are used to train the convolutional neural net.  Specifically, the training set\textemdash composed of twenty-one hydrocarbon molecules\textemdash follows: 2.01: `Ethane', 3.01: `Propane', 4.01: `Butane', 4.02: `2-Methylpropane', 5.01: `Pentane', 5.02: `2-Methylbutane', 5.03: `2,2-Dimethylpropane', 6.01: `Hexane', 6.02: `2-Methylpentane', 6.03: `3-Methylpentane', 6.04: `2,2-Dimethylbutane', 6.05: `2,3-Dimethylbutane', 7.01: `Heptane', 7.02: `3-Methylhexane', 7.03: `2-Methylhexane', 7.04: `2,2-Dimethylpentane', 7.05: `2,3-Dimethylpentane', 7.06: `2,4-Dimethylpentane', 7.07: `3,3-Dimethylpentane', 7.08: `3-Ethylpentane', 7.09: `2,2,3-Trimethylbutane'.

\vspace{2mm}
\noindent \textit{Testing Data.} All isomers of octane isomers as well as nonane, decane, undecane, dodecane, tridecane, tetradecane, and pentadecane are used to test the trained neural net.  Specifically, the testing set follows:  8.01: `Octane', 8.02: `2-Methylheptane', 8.03: `3-Methylheptane', 8.04: `4-Methylheptane', 8.05: `2,2-Dimethylhexane', 8.06: `2,3-Dimethylhexane', 8.07: `2,4-Dimethylhexane', 8.08: `2,5-Dimethylhexane', 8.09: `3,3-Dimethylhexane', 8.1: `3,4-Dimethylhexane', 8.11: `3-Ethylhexane', 8.12: `2,2,3-Trimethylpentane', 8.13: `2,2,4-Trimethylpentane', 8.14: `2,3,3-Trimethylpentane', 8.15: `2,3,4-Trimethylpentane', 8.16: `3-Ethyl-2-Methylpentane', 8.17: `3-Ethyl-3-Methylpentane', 8.18: `2,2,4,4-Tetramethylbutane', 9.01: `Nonane', 10.01: `Decane', 11.01: `Undecane', 12.01: `Dodecane', 13.01: `Tridecane', 14.01: `Tetradecane', 15.01: `Pentadecane'.

\vspace{2mm}
\noindent \textit{CNN Specifics.} The convolutional neural network is composed of an input layer, five additional dense layers, and an output layer.  The input layer consists of partition functions and the Hartree-Fock correlation temperature as specified in the \textit{Model Inputs} section, and the output layer is a dense layer consisting of the $\Delta$ value described in the \textit{Model Outputs} section.  The additional dense layers have 503, 240, 100, 50, and 20 nodes, respectively.  All dense nodes are initialized via the $he\_uniform$ kernel initializer with a $relu$ activation function.   For the training of the convolutional net, loss is measured via mean absolute error, and the $adam$ optimizer is implemented for $30,000$ epochs. This convolutional neural network is implemented using Keras\textemdash Python's deep learning API \cite{chollet2015keras}.

\vspace{3mm}
\color{black}
\noindent\textbf{Computational Scaling}

\noindent For the testing set, scaling is dominated by the determination of the geminal energies, which are obtained via the diagonalization of the two-electron reduced Hamiltonian, a computation that scales as $O[r^6]$ where $r$ is the number of orbitals in the active space.  Thus, for a given molecule in the testing set, computational expense for prediction of molecular energies scales as $O[r^6]$.  The computational expense of the training set is dominated by the determination of the reference CASSCF or CCSD energies necessary to obtain the reference correlation temperature\textemdash which are known to scale approximately as $O[N!]$ and $O[N^6]$, respectively, for a given molecule where $N$ for CASSCF is the number of active electrons and $N$ for CCSD is the number of total electrons.

\color{black}

\vspace{2cm}
\newpage
\section{\color{black}TOC Graphic \color{black}}
\begin{figure}[h!]
    \hspace{0cm}
    \centering
    \includegraphics[width=3.25in]{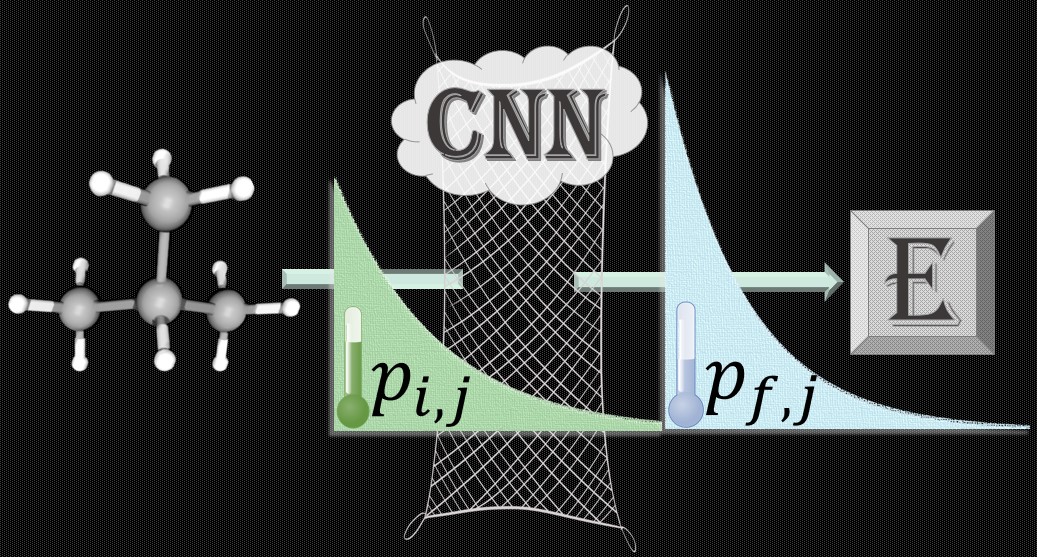}
\end{figure}

\end{document}